# Ontology Based Query Expansion Using Word Sense Disambiguation

M. Barathi
Department of Computer Applications
S. M. K. Fomra Institute of Technology
Chennai, India .

S.Valli
Department of Computer Science and Engineering
Anna University
Chennai, India .

*Abstract -* The existing information retrieval techniques do not consider the context of the keywords present in the user's queries. Therefore, the search engines sometimes do not provide sufficient information to the users. New methods based on the semantics of user keywords must be developed to search in the vast web space without incurring loss of information. The semantic based information retrieval techniques need to understand the meaning of the concepts in the user queries. This will improve the precision-recall of the search results. Therefore, this approach focuses on the concept based semantic information retrieval. This work is based on Word sense disambiguation, thesaurus WordNet and ontology of any domain for retrieving information in order to capture the context of particular concept(s) and discover semantic relationships between them.

*Index terms – Word Sense Disambiguation, Semantic Information Retrieval, Clustering, Ontology.*

## I. INTRODUCTION

Search engines have become the most helpful tools for obtaining useful information ever since the development of the World Wide Web. But, the search engines sometimes fail to cater to the users need. The huge volume of information accessible over networks makes it difficult for the user to find exact information needed. Numerous information retrieval techniques have been developed based on keywords. These techniques use keyword list to describe the content of the information without addressing anything about the semantic relationships of the keywords. As a result, understanding the meaning of the keyword becomes difficult [1]-[4]. Synonym and polysemy are two prominent issues. A synonym is a word which means the same as another word. For instance the word *animal* is a synonym of a living organism. A polysemy is a word with multiple, related meanings. For example, the word *cell* can be used to refer to a small room in one context and the basic structural and functional unit of an organism in another context [1], [3]-[4]. In WordNet, the word *cell* has multiple meaning as shown in Figure 1. So, *cell* is a polysemy word.

| Key word | Sense | Noun | Synonyms |
|---|---|---|---|
| Cell | Any small compartment | ✓ | The cells of a honeycomb |
| | biology | ✓ | The basic structural and functional unit of all organisms. |
| | Electric cell | ✓ | A device that delivers an electric current as the result of a chemical reaction. |
| | Cadre | ✓ | A small unit serving as part of or as the nucleus of a larger political movement. |
| | Cellular Telephone, Cellular Phone, Cell Phone, Mobile phone | ✓ | A hand-held mobile radiotelephone for use in an area divided into small sections, each with its own short-range transmitter/receiver. |
| | Cubicle | ✓ | small room in which a monk or nun lives |

Figure 1. Multiple Meaning for the word *"cell"*

In semantic-based information retrieval techniques searching is performed by interpreting the meanings of the keywords (i.e., semantic). The system which retrieves information based on the semantics of the keyword attains higher precision than the one which is based on the keyword. Domain ontologies are used as knowledge base to understand the meanings of the concepts.

The semantic-based information retrieval techniques, search by interpreting the meanings of the keywords by sensing the word using the thesaurus WordNet. It is often difficult for ordinary users to use information retrieval systems based on these commonly used keyword based techniques. So Tin Berners-Lee introduced the idea of a semantic web, where machine readable Semantic knowledge is attached to all information. The Semantic knowledge attached to the information is united by means of ontologies, i.e., the concepts attached to the information are mapped into these ontologies. Ontology is "a formal, explicit specification of a shared conceptualization" [5]. Ontology is arranged in a lattice or taxonomy of concepts in classes and subclasses (cancer, inflammatory, clumps, metastasis) as shown in Figure 2. Each concept is typically associated with various properties describing its features and attributes as well as various restrictions on them. Ontology together with asset of concrete instances (also called individuals) of the class constitutes a knowledge base. The semantics of keywords are identified through the relationships between keywords by performing semantic similarity on them[6],[1],[7]-[9],[2],[10].

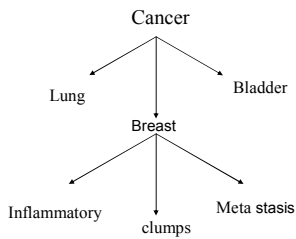

Figure 2. A Sample Domain Ontology

In our proposed work we use Word sense disambiguation to disambiguate several meaning for a word. Word Sense Disambiguation (WSD) is defined as the process of enumerating the sense of a word. The frequency of the keywords occurring in the web pages is calculated and they are ranked using the traditional weighting scheme [19] tfidf values and stored in the database. To enrich the user query for efficient retrieval of web pages, the user query is matched with the set of k-cores [11] which is constructed using tfidf values. The user query is enriched with the most relevant k-core using WSD and passed to the search engine for the retrieval of the relevant web pages. In order to refine the web search using ontology, both the k-cores and the ontology of a medical domain is used to enrich the user query for more efficient retrieval of web pages. The relevant k-cores are matched with the ontology of a particular domain to extract the concepts based on the similarity measure. The concepts are extracted by the concept extractor based on the most relevant k-cores. The most relevant concepts along with the ranked k-cores are presented to the user to choose the best concept for expansion. This is supposed to the best as the user himself disambiguates. The user query is enriched with the selected concept and passed to the search engine for efficient retrieval of relevant web pages.

K-core is a kind of keyword cluster. K-cores are the core words of a concept or theme. Each k-core is a representative of the semantic context. K-core is a set of keywords obtained per topic in a set of web pages. In this approach set of four keywords form a k-core in order to perform meaningful experiments. The keywords are clustered (i.e., k-core) and ranked according to the keyword frequency count. For example consider the topic as cancer. The best 4 cores are given in Table 1.

Table 1. A sample k-core for the topic *cancer*

| K-Core |
| --- |
| Cancer, Oncology, Oncogene, Meta Stasis |
| Disease, Cancer clumps, Treatment |
| Cancer, Breast, Lump, Leukemia |

The goal of this work is to choose the best concept and expand the user query for efficient retrieval of information to satisfy the user needs and expectation. The rest of the paper is as follows. In section 2, the existing works are highlighted. In section 3 our proposed methodology is explained. Section 4 presents the experimental results obtained and section 5 concludes the work.

## II. EXISTING WORKS

In Seamless searching of Numeric and Textual Resources project [12] the author use a customized dictionary to disambiguate the concepts used for querying. However our system uses a general-purpose thesaurus, WordNet, and the context of the user keywords. CUPID [13] and onto builder [14] identify and analyze the factors that affect the effectiveness of algorithms for automatic semantic reconciliation. Our system uses a set of k-core and WSD to disambiguate the concepts and the ontology of a particular domain to enrich the user query for more efficient retrieval of information. GLUE[15] studies the probability of matching two concepts by analyzing the available ontologies, using a relaxation labeling method[16]; however, this approach is not very flexible or / adaptable because it analyzes all of the ontology concepts, while we use an approach based on word sense disambiguation to disambiguate the senses and expand the user query with the best concept. The internet searches can be much more focused so that only relevant web pages are retrieved.

## III. CONCEPT BASED INFORMATION RETRIEVAL FRAMEWORK

The proposed system refines the web search for efficient retrieval of web pages. Only web pages specific to the context are retrieved. The set of web pages of a particular domain are retrieved from the web by web searches. Hill-climbing algorithm is used to "mine" a set of web pages for finding k-cores. Stop words are removed and the core keywords are extracted.

The keywords (i.e. k-core) are clustered based on tfidf values. Each k-core is a representative of the semantic context. Here, k is the size of the cluster. It may be 3 or 4 k-core. Using the frequency count of keywords, the web searches can be much more focused so that only relevant web pages are retrieved. This process is shown in Figure 3.

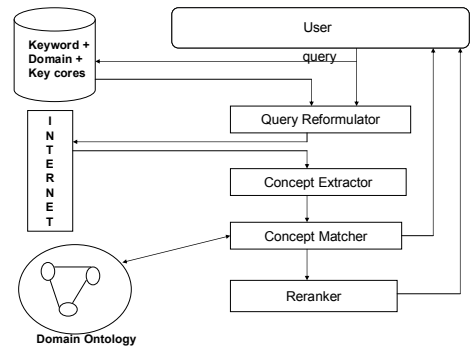

Figure 4. Concept Based Semantic Information Retrieval Framework

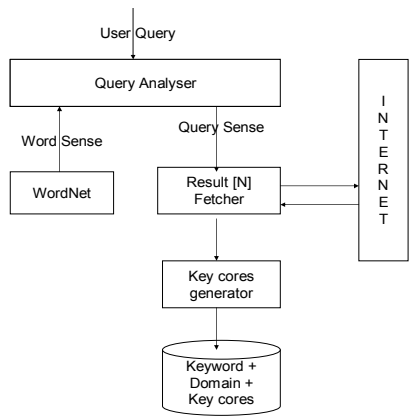

Figure 3. Block Diagram of Web page retrieval and Keycore Generation Module

The user query along with the set of k-cores is refined by using WSD to disambiguate the senses for efficient retrieval of web pages. The refined user query is passed to the search engine for retrieving the relevant web pages. For further refinement of the user query the K-cores and the ontology of the particular domain are used for retrieving relevant web pages. The concepts are extracted by the concept extractor based on the most relevant k-cores. The most relevant concepts along with the ranked k-cores are presented to the user to choose the best concept for expansion. This is supposed to be the best, as the user himself disambiguates. The user query is enriched with the selected concept and passed to the search engine for efficient retrieval of relevant web pages as shown in Figure 4. This framework consists of components namely the Query reformulator, Concept Extractor, Matcher and Reranker.

A. *Query Reformulator*

The query reformulator expands the query using relationship such as synonym [6],[1],[8],[5], semantic neighborhood [2], hyponym [6],[2],[17] (i.e. Is-A relationship) and Meronym (i.e. Part-of) [6],[2] using distance based approach [18],[8],[9]. Then the query is rewritten with these expanded terms and passed to the Concept Matcher.

B. *Concept Extractor and Matcher*

The concept matcher matches the relevant k-cores with the ontology to extracts the concepts based on the similarity measure. The concepts are extracted by the concept extractor based on the most relevant k-cores.

C. *Re-ranker*

The Reranker sorts the document according to the relevance of the user's queries. Documents that are related to the user's query are retrieved and ranked according to their importance. The relevance between the documents and frequency count of the keywords are measured. The relevance of the documents is computed using the traditional weighting scheme [19] given in equation (1). The tf in equation (1) refers term frequency, N is the total no. of documents, df is the document frequency and tfidf is the term frequency inverted document frequency.

$$tfidf = tf * \log(N/df) \quad (1)$$

D. *Concept Similarity*

The similarity is calculated by measuring semantic similarity of concepts and their relationships. The concepts similarity is measured by calculating the distance between them [18], [8], [9]. The distance is calculated between different concepts from their position in the concepts hierarchy. The position of a concept in a hierarchy is defined [8] using equation (2), where 'k' is a predefined factor larger than 'l' and l(n) is the depth of the node 'n' in hierarchy.

$$\text{Milestone}(n) = \frac{\frac{1}{2}}{K^{l(n)}} \quad (2)$$

For the root of a hierarchy, l (root) is zero. For any two concepts c1 and c2 in the hierarchy having closest common parent (ccp), the distance $d_c$ between two concepts and their ccp is calculated using equations (3) and (4).

$$d_c(c1,c2) = d_c(c1,ccp) + d_c(c2,ccp) \quad (3)$$

$$d_c(c1,cpp) = \text{milestone}(ccp) - \text{milestone}(c1) \quad (4)$$

Thus, the similarity $sim_c$ between the two concepts $c_1$ and $c_2$ is calculated using equation (5)

$$sim_c(c1,c2) = 1 - d_c(c1,c2) \quad (5)$$

If the concept c1 and concept c2 are synonym or acronym of each other, the distance will be 0, i.e. the similarity between these two concepts will be 1. Synonym and acronym relation between concepts are treated at the same level.

*E. Relations Similarity*

The similarity $Sim_r$ between any two relations r1 and r2 is given by equation (6)

$$Sim_r(r1,r2) = 1 - d_r(r1, r2) \quad (6)$$

The distance between two relations is also calculated by their respective positions in the relation hierarchy.

*F. Web Page Retrieval*

The web crawler receives the user query of any domain from the user interface and downloads the web pages corresponding to that domain from the web. Then it opens the URL connection and reads the content of the web page and stores it in the text file. If the web page contains another URL, it adds the URL to the end of the list of URLs to be crawled. It repeats the process until all the URLs in the list are crawled as shown in Figure 5. The field "IsCrawled" in Figure 5 represents that the particular URL has been crawled by setting the value as "t".

| CRAWLTABLE ||||
|---|---|---|
| Serial | URL Address | IsCrawled |
| 123 | http://localhost:8080/cancer.html | t |
| 124 | http://localhost:8080/cancertypes.html | t |
| 125 | http://localhost:8080/leukemia.html | t |
| 126 | http://localhost:8080/causes.html | t |
| 127 | http://localhost:8080/oncology.html | t |
| 128 | http://localhost:8080/oncogenes.html | t |

Figure 5. Sample list of URLs crawled from the web

*G. Refining Web Search Using Ontology*

The k-cores and the ontology of the particular domain are used in enhancing the user query for more efficient retrieval of web pages. The first step matches the user query with the set of k-cores using WordNet to disambiguate the senses. Then the relevant k-cores are matched with the ontology of the particular domain based on the similarity measure. The concepts are extracted by the concept extractor based on the most relevant k-cores. The most relevant concepts along with the ranked k-cores are presented to the user to choose the best concept for expansion. This is supposed to the best as the user himself disambiguates. The user query is enriched with the selected concept and passed to the search engine for efficient retrieval of relevant web pages. The algorithm for refining the web search using Ontology is given in Figure 7.

---

**Input :** User query, Domain ontology and K-cores
**Output :** Set of relevant web pages
**Algorithm**

1. The user query and the set of k-cores are disambiguated using thesaurus WordNet.
2. The set of relevant k-cores is matched with the concepts in the domain ontology to extract the relevant concepts based on the similarity measure.
3. The set of relevant concepts along with the k-cores are presented as options to the user to disambiguate the senses.
4. When the user selects the specific concepts he wants, the user query is enriched with that concept.
5. The enriched query is passed to the web searcher and it displays the relevant results to the user.
6. End

---

Figure 7. Algorithm for Refining the web search using Ontology

## IV EXPERIMENTAL RESULTS

This work is implemented using Java and the medical ontology is used as the domain. In this study, around 1500 web pages have been crawled using Google search engine and stored as text document. Preprocessing is done on each document which is the text file as shown in Figure 9.

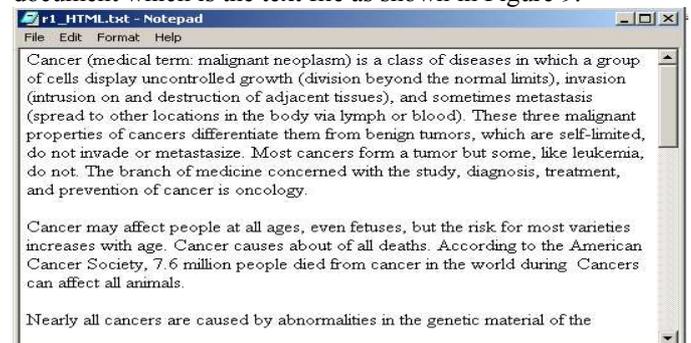

Figure 9. Text File

Then the frequency count for each term is calculated as shown in Figure 10 and term frequency inverted document frequency for the terms are calculated and the ranked list is shown in Figure 11.

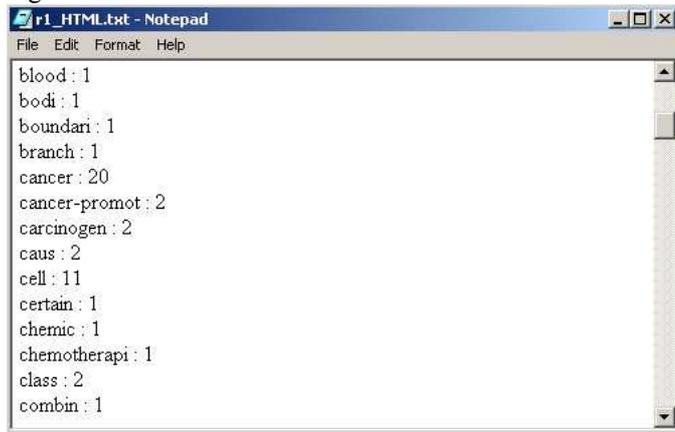

Figure 10 Keyword frequency list

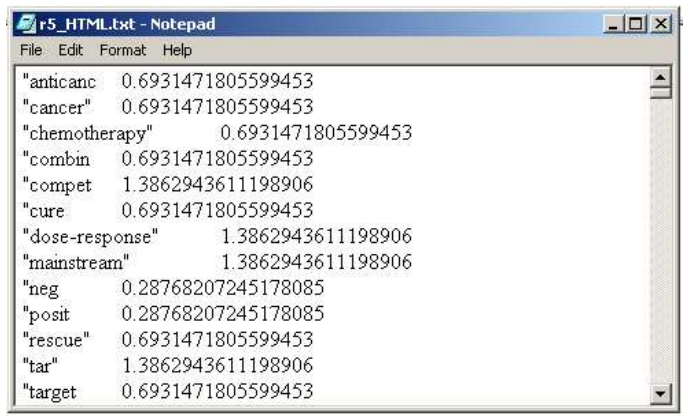

Figure 11. List of term frequency inverted document frequency for the keywords

Using this ranked list, set of k-cores are constructed. When the query is entered, the user query is matched with the set of k-cores. WordNet shows different senses for the word *cancer* such as cancer zodiac, cancer horoscope, type of cancer etc. To disambiguate these senses, user selects the best synonyms he is looking for. This enhanced query is passed to the search engine for retrieving relevant web pages. An example of user query and the enhanced query is shown in Figure 12.

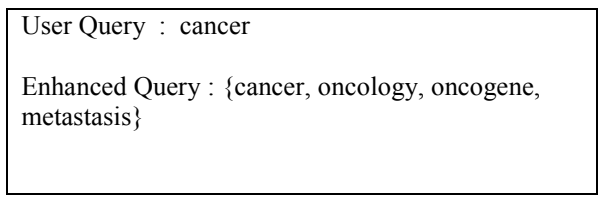

User Query : cancer

Enhanced Query : {cancer, oncology, oncogene, metastasis}

Figure 12. A sample of original query and Enhanced query

In order to refine the web search using ontology, both the k-cores and the ontology of breast cancer is used to enhance the user query for more efficient retrieval of web pages. A piece of breast cancer ontology is shown in Figure 13. The relevant k-cores are matched with the ontology breast cancer to extract the concepts based on the similarity measure. Then the user query is enhanced with the selected concept and passed to the search engine for efficient retrieval of relevant web pages.

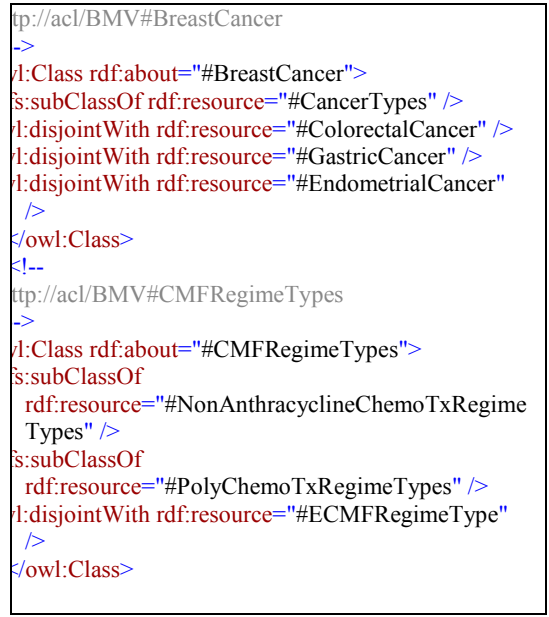

Figure 13 A piece of ontology for Breast cancer

Using recall and precision, the effectiveness of an Information retrieval system is evaluated. The most often used common evaluation measures are precision and recall as given in equation (7) and (8). Precision measures the proportion of retrieved documents that are relevant, and recall measures the proportion of relevant documents that have been retrieved. They are defined as follows

$$\text{Precision} = \frac{\text{Retrieved relevant documents}}{\text{Retrieved documents}} \quad (7)$$

$$\text{Recall} = \frac{\text{Retrieved relevant documents}}{\text{All relevant documents}} \quad (8)$$

The precision is measured at a number of standard recall values (i.e. recall assumes the following values 0.1, 0.2, 0.3, 0.4, 0.5, 0.6, 0.7, 0.8, 0.9, 1.0). These measurements result in a set of recall precision figures. These figures are presented in Figure 14.

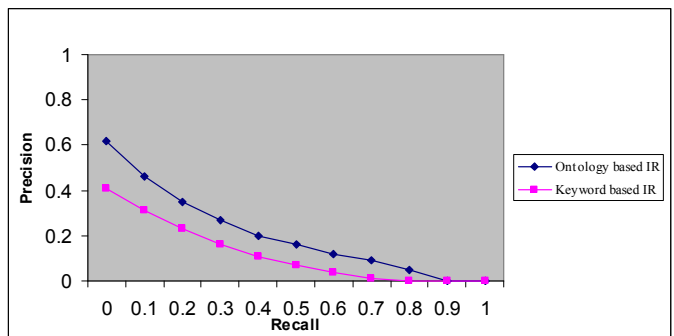

Figure 14. Precision –Recall graph for Ontology based IR and Keyword based IR

V CONCLUSION

This paper addresses an Ontology based Query expansion to improve the precision-recall of the search results by concentrating on the context of concept(s). The relevant k-cores are matched with the ontology of medical domain to extract the concepts based on the similarity measure. The most relevant concepts along with the ranked k-cores are presented to the user. The user query is enriched with the selected concept and passed to the search engine for efficient retrieval of relevant web pages. The future work would focus on the automatic selection of concepts i.e. Intelligent WSD suitable for user's information need.